\documentclass[a4paper]{jpconf}
\usepackage{graphicx}
\begin{document}
\title{A simple method for the Kramers-Kronig analysis of 
reflectance spectra measured with diamond anvil cell }

\author{H. Okamura}

\address{Department of Physics, Graduate School of Science, 
Kobe University, Kobe 657-8501, Japan}

\ead{okamura@kobe-u.ac.jp}

\begin{abstract}
When the optical reflectance spectrum of a sample under 
high pressure is studied with a diamond anvil cell, it is 
measured at a sample/diamond interface.   Due to the large 
refractive index of diamond, the resulting reflectance 
$R_{\rm d}(\omega)$ may substantially differ from that measured 
in vacuum.   
To obtain optical constants from $R_{\rm d}(\omega)$, therefore, 
the usual Kramers-Kronig (KK) analysis cannot be straightforwardly 
applied, and either a spectral fitting or a modified KK transform 
has been used.   
Here we describe an alternative method to perform KK 
analysis on $R_{\rm d}(\omega)$.    
This method relies on the usual KK transform with an 
appropriate cutoff and extrapolation to $R_{\rm d}(\omega)$, 
and may offer a simpler approach to obtain infrared 
conductivity from measured $R_{\rm d}(\omega)$.

\end{abstract}

\section{Introduction}
Infrared (IR) spectroscopy has been a powerful tool 
to study the microscopic carrier dynamics and electronic 
structures in strongly correlated electron materials, 
such as rare earth ($f$ electron), transition metal 
($d$ electron), and organic ($p$ electron) 
compounds \cite{basov}.   
The IR spectroscopy technique has been also performed 
under high pressure using a diamond anvil cell (DAC) 
[2-13] since the strongly correlated materials show many 
interesting physical properties under high pressure.      
In a DAC, a pair of diamond anvils and a thin metal 
gasket are used to seal a sample and a pressure 
transmitting medium \cite{airapt}.  
A typical diameter of the diamond surface is 0.8~mm 
to reach a pressure of 
10~GPa, and 0.6~mm to reach 20~GPa.   Therefore the 
sample in this experiment should have dimensions 
of the order of 100~$\mu$m.   To perform an infrared 
(IR) reflectance study on such a small sample under 
the restricted sample space in a DAC, synchrotron radiation 
(SR) has been used as a 
bright source of both far and mid-infrared.   In fact, 
high pressure IR spectroscopy with DAC is currently 
one of the major applications of 
IR-SR [4-7,9-13].

With a DAC, the reflectance is measured between the 
sample/diamond interface, in contrast to the usual case 
of sample/vacuum or sample/air interface.   
The normal-incidence reflectance of a sample 
relative to a transparent medium of (real) refractive 
index $n_0$ is given by the Fresnel's formula as 
\cite{wooten,dressel}: 
\begin{equation}
R(\omega)=\frac{(n-n_0)^2+k^2}{(n+n_0)^2+k^2}.  
\end{equation}
Here, $\hat{n} = n + ik$ is the complex refractive index of 
the sample, and $n_0$=2.4 for diamond and 1.0 for vacuum.   
Hereafter, we denote $R(\omega)$ at sample/diamond interface 
as $R_{\rm d}(\omega)$, and that at sample/vacuum interface 
as $R_0(\omega)$.   From Eq.~(1), it is easily seen that 
$R_{\rm d}(\omega)$ of a sample measured in DAC may be 
substantially different from $R_0(\omega)$.   

The purpose of this study is to consider the 
Kramers-Kronig (KK) analysis of $R_{\rm d}(\omega)$ data 
measured in DAC.   KK analysis has been widely used 
to derive optical constants such as the refractive 
index, dielectric function and optical conductivity 
from a measured $R_0(\omega)$ spectrum \cite{wooten,dressel}.  
However, due to 
the difference between $R_0(\omega)$ and $R_{\rm d}(\omega)$ 
discussed above, the usual KK analysis method cannot be 
straightforwardly applied to $R_{\rm d}(\omega)$ \cite{kk}.   
To derive optical constants from $R_{\rm d}(\omega)$, 
therefore, previous high pressure IR studies used either 
a Drude-Lorentz spectral 
fitting \cite{syassen,ybs,airapt,sku,elettra,uvsor,anka} 
or a modified KK transform 
\cite{organic,organic2,degiorgi,lupi}.
In this work, we propose a different method, which relies 
on the usual KK transform with an appropriate cutoff to the 
$R_{\rm d}(\omega)$, as an alternative approach to obtain 
the infrared $\sigma(\omega)$ from $R_{\rm d}(\omega)$.    
The validity of the proposed method is demonstrated with 
actually measured reflectance data of PrRu$_4$P$_{12}$.

\section{Kramers-Kronig analysis of reflectance spectra}   
The complex reflectivity of the electric field, $\hat{r}$, 
is expressed as \cite{wooten,dressel} 
%%%%%%%%%%%%%%%%%%%
\begin{equation}
\hat{r}(\omega)= \frac{n_0 - \hat{n}(\omega)}{n_0 + \hat{n}(\omega)}
= r(\omega) e^{i \theta(\omega)}. 
\end{equation}
%%%%%%%%%%%%%%%%%%%%
Here $r(\omega)$ is the square root of the 
reflectance $R(\omega)$, which is actually measured in 
experiments.
Then the real and 
imaginary parts of $\hat{n}$ can be expressed in terms 
of $r$ and $\theta$ as 
%%%%%%%%%%%%%%%%%%
\begin{equation} 
n= \frac{1-r^2}{1+r^2+2r\cos{\theta}}\cdot n_0   
\end{equation}
%%%%%%%%%%%%%%%%%%%
and
%%%%%%%%%%%%%%%%%%%%%%%%
\begin{equation}
k= \frac{-2r\sin{\theta}}{1+r^2+2r\cos{\theta}}\cdot n_0, 
\end{equation}
%%%%%%%%%%%%%%%%%%%%%%%%
respectively.   Therefore, if $\theta(\omega)$ can be 
derived from measured $r(\omega)$ with KK analysis 
even for the sample/diamond reflection case, $n(\omega)$ and 
$k(\omega)$ can also be derived simply by setting $n_0$=2.4 
in Eqs.~(2) and (3).  Then, the imaginary part of the complex 
dielectric function is given as 
$\epsilon_2=2n(\omega)k(\omega)$, and the 
optical conductivity is given as 
$\sigma(\omega)=\frac{\omega}{4\pi}\epsilon_2(\omega)$ 
\cite{wooten,dressel}.

In performing KK analysis on reflectance data, 
usually the logarithm of $\hat{r}$, namely 
%%%%%%%%%%%%%%%%%%%%%%%%%%%%
\begin{equation}
\ln{\hat{r}(\omega)} = \ln{r(\omega)} + i \theta(\omega)
\end{equation}
%%%%%%%%%%%%%%%%%%%%%%%%%%%%%
is regarded as a complex response function.  In the 
case of sample/vacuum reflection, the KK relation 
between $\ln{r(\omega)}$ and $\theta(\omega)$ is 
expressed as \cite{wooten,dressel} 
%%%%%%%%%%%%%%%%%%%%%%%%%%%%%
\begin{equation}
\theta(\omega) = - \frac{2\omega}{\pi} P \int_0^\infty 
 \frac{\ln{r(\omega^\prime)}}{\omega^{\prime 2} - \omega^2}
d\omega^\prime.  
\end{equation}
%%%%%%%%%%%%%%%%%%%%%%%%%%%%%%
Here, $P$ denotes the principal value.  
In deriving this relation, it is required that 
$\ln{\hat{r}(\hat{\omega})}$ has no poles 
in the upper complex $\hat{\omega}$ plane when 
$|\omega |$ is finite.    
This is correct since $r(\omega) \rightarrow 0$ 
only when $\omega \rightarrow \infty$ in the case 
of $n_0$=1.   
However, when $n_0 > 1$ as in the case of sample/diamond 
interface, $\hat{n}=n_0$ may be satisfied at some point 
on the upper imaginary axis \cite{kk}.   
This point is denoted as $\hat{\omega}=i\beta$, 
where $\beta$ is a real, positive and finite number.  
When $\hat{n}=n_0$, $\hat{r}=0$ from Eq.~(2) and 
$\ln{\hat{r}}$ therefore has a pole at 
$\hat{\omega}=i\beta$. 
Accordingly, the KK relation in this case must be 
modified to \cite{kk} 
%%%%%%%%%%%%%%%%%%%%%%%%%%%%%%%%%
\begin{equation}
\theta(\omega) = - \frac{2\omega}{\pi} P \int_0^\infty 
 \frac{{\rm ln} r(\omega^\prime)}{\omega^{\prime 2} - \omega^2}
d\omega^\prime + 
[\pi - 2 \arctan{(\beta/\omega)}].   
\end{equation}
%%%%%%%%%%%%%%%%%%%%%%%%%%%%%%%%%
Namely, the presence of a medium with $n_0 > 1$ brings 
an extra phase shift, indicated by the 
square bracket in Eq.~(7), into the KK relation. 
Note that the extra phase shift is a decreasing function 
of $\beta/\omega$, and that the 
original KK relation of Eq.~(6) is recovered when 
$\beta/\omega \rightarrow \infty$ \cite{kk}.  
Detailed theoretical considerations on the extra phase 
shift in various situations have been reported \cite{kk2,kk3}.  
In the case of actual experimental studies, however, 
the precise value of $\beta$ may not be known.    
Accordingly, the value of $\beta$ has been estimated 
from experimental $R_{\rm d}(\omega)$ data by use of a 
combination of DL fitting and the modified KK 
transform \cite{organic,organic2,degiorgi,lupi}.  
In this method, one uses Eq.~(7) with 
$r(\omega)=\sqrt{R_{\rm d}(\omega)}$ and looks for a value 
of $\beta$ that well reproduces the 
$\sigma(\omega)$ given by a DL fitting of $R_{\rm d}(\omega)$.

Note that, on the other hand, 
if the frequency range of interest 
is lower than the value of $\beta$, effects 
of the extra phase may be only minor, and the 
usual KK transform of Eq.~(6), combined with the 
use of $n_0$=2.4 in Eqs.~(3) and (4), might give 
sufficiently accurate values of optical constants.  
We will examine the validity of such a procedure 
in the next section.

\section{Simulation with measured reflectance spectra}  
Here, we use $R_0(\omega)$ spectra actually measured on 
PrRu$_4$P$_{12}$ \cite{matunami}.   
This compound is well known for showing a metal-to-insulator 
transition at about 60~K \cite{sekine}, and a clear 
energy gap in $\sigma(\omega)$ was observed in our 
previous work \cite{matunami}.  Here we use $R_0(\omega)$ 
at 60~K (metal) and 9~K (insulator) measured over a wide 
photon energy range of 0.008-30~eV \cite{matunami}, which 
are shown by the blue curve in Figs.~1(a) and 2(a).    
%%%%
%%%%%%%%%%%%%%%%%  FIG. 1  %%%%%%%%%%%%%
\begin{figure}
\begin{center}
\includegraphics[width=22pc]{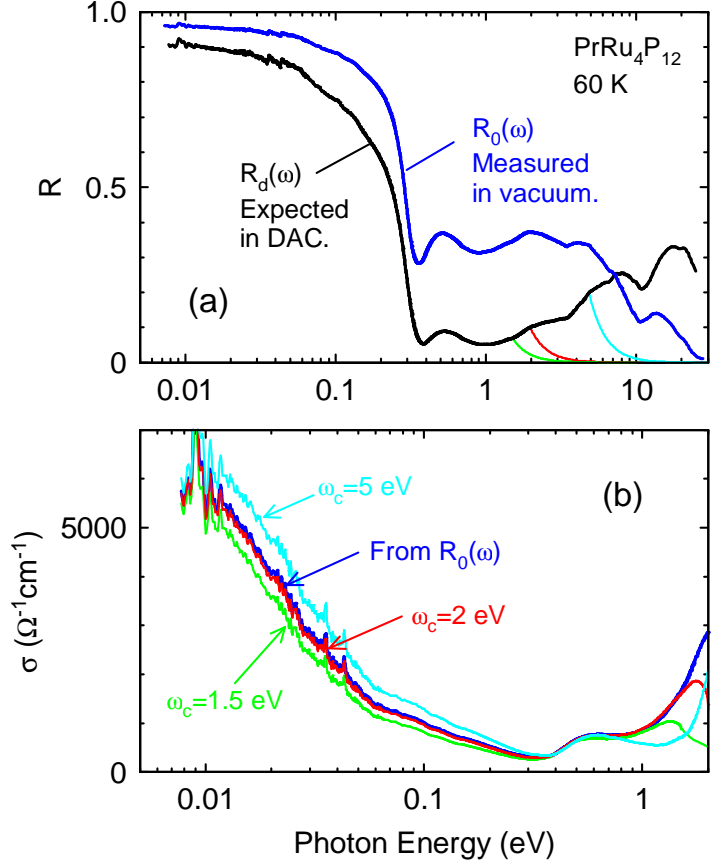}
\end{center}
\caption{(a) $R_0(\omega)$ is the reflectance spectrum of 
PrRu$_4$P$_{12}$ measured at 60~K in vacuum \cite{matunami}, 
and $R_{\rm d}(\omega)$ is that expected in a DAC calculated 
from $R_0(\omega)$ as described in the text.  The green, red, 
and light blue curves are $\omega^{-4}$ extrapolations with 
cutoff energies of $\omega_c$=1.5, 2, and 5~eV, respectively.  
(b) The optical conductivity ($\sigma$) obtained with KK 
analysis of $R_0(\omega)$ is compared with those obtained 
with KK analysis of $R_{\rm d}(\omega)$ with a cutoff at 
$\omega_c$=1.5, 2, and 5~eV, and $\omega^{-4}$ extrapolations 
above them.   
Below 1.5~eV, $\sigma(\omega)$ obtained from 
$R_{\rm d}(\omega)$ with $\omega_c$=2~eV agrees very well 
with that obtained from $R_0(\omega)$.  
}
\end{figure}
%%%%%%%%%%%%%%%%%%%%%%%%%%%%%%%%%%%%%%%%%%
%%%%
The procedure is the following.    
\begin{itemize}
 \item[(i)]The full $R_0(\omega)$ spectrum is KK 
analyzed with Eq.~(6) to obtain $n(\omega)$, 
$k(\omega)$ and $\sigma(\omega)$.    

\item[(ii)]The above $n(\omega)$ and $k(\omega)$ are 
substituted into Eq.~(1) with $n_0$=2.4 to derive 
$R_{\rm d}(\omega)$ that is {\it expected} in a DAC.  

\item[(iii)] The $R_{\rm d}(\omega)$ obtained above is 
used with the usual KK transform of Eq.~(6) and 
$n_0$=2.4 in Eqs.~(3) and (4), to obtain 
$n(\omega)$, $k(\omega)$ and $\sigma(\omega)$.   
Before this is done, an appropriate cutoff and 
extrapolation are made to the $R_{\rm d}(\omega)$, 
as described in detail below.  
\end{itemize}
If the KK analysis on 
$R_{\rm d}(\omega)$ works properly, the resulting 
$\sigma(\omega)$ from (iii) should well agree with 
that given by $R_0(\omega)$ and the usual KK analysis.

We first examine the 60~K data.  The expected 
$R_{\rm d}(\omega)$ at 60~K obtained by (i) and 
(ii) is indicated by the black curve in Fig.~1(a).    
In carrying out the integration in Eq.~(6), 
the $R_0(\omega)$ spectrum were extrapolated below 
0.008~eV and above 30~eV with the Hagen-Rubens 
and $\omega^{-4}$ functions, respectively.  
It is seen in Fig.~1(a) that $R_{\rm d}(\omega)$ shows 
very high values above about 4~eV.   This physically 
unrealistic feature resulted from the unphysical 
assumption of constant $n_0$=2.4 in the entire spectral 
range.  
In reality, of course, the refractive index of 
diamond cannot be constant and real near and above 
the band gap, where it shows strong light absorption.  
In addition, when $\omega \rightarrow \infty$, 
$n(\omega) \rightarrow 1$ and $k(\omega) \rightarrow 0$, 
and therefore $R_{\rm d}(\omega) \rightarrow 0$ must hold 
just like any other material.   
Accordingly, before performing KK transform in (iii), 
$R_{\rm d}(\omega)$ in Fig.~1(a) was cut off at some 
energy $\omega_c$, and then it was extrapolated with 
$\omega^{-4}$ function.  
Several different values of $\omega_c$ were tried, as 
shown in Fig.~1(a).   
For each value of $\omega_c$, the usual KK transform 
of Eq.~(6) was made to get $\theta(\omega)$, which was 
then used to derive $n(\omega)$ and $k(\omega)$ with 
$n_0$=2.4 in Eqs.~(3) and (4), and to finally obtain 
$\sigma(\omega)$.

Figure~1(b) shows the $\sigma(\omega)$ spectra 
obtained as described above, with different values of 
$\omega_c$ which are also indicated in Fig.~1(a).  
It is seen that the obtained $\sigma(\omega)$ spectra 
strongly depend on $\omega_c$. 
With $\omega_c$=2.0~eV, the resulting $\sigma(\omega)$ 
below 1.5~eV agrees very well with that derived from 
the original $R_0(\omega)$.     
(Actually, $\omega_c$=2.2~eV gives the best agreement, 
but $\omega_c$=2.0~eV data is shown instead.  This is 
because the 2.2~eV data almost completely overlaps with 
that from $R_0(\omega)$, making it difficult to 
distinguish them in the figure.)  
The result for the 9~K data, where the sample is an insulator 
(semiconductor), is also shown in Fig.~2.  A good agreement 
is again observed between the $\sigma(\omega)$ derived from 
the full $R_0(\omega)$ and that from $R_{\rm d}(\omega)$ 
with $\omega_c$=2~eV.   The spectral range of their good 
agreement is below 1.5~eV, which is similar to the 
case of 60~K data discussed above.  
Note also that both 9~K and 60~K data show good 
agreement with the common value of $\omega_c$= 2~eV.   
These results show that, for any spectral change in 
$R_{\rm d}(\omega)$ (either temperature- or pressure-induced) 
below 1.5~eV, the corresponding $\sigma(\omega)$ can be 
obtained by the present method.  
In actual high pressure studies of strongly correlated 
materials with DAC [2-13], $R_{\rm d}(\omega)$ is usually 
measured below 1-2~eV.  
Hence, above the high energy limit of $R_{\rm d}(\omega)$ 
measurement, the $R_{\rm d}(\omega)$ expected from 
$R_0(\omega)$ can be connected to the measured 
$R_{\rm d}(\omega)$, with the cutoff and extrapolation 
discussed above.   Then the connected $R_{\rm d}(\omega)$ 
may be KK 
transformed to obtain $\sigma(\omega)$, as discussed above.    
An obvious condition required for this method to work 
properly is that the pressure- and temperature-induced 
changes of $R_{\rm d}(\omega)$ should be limited below 
certain energy, which is 1.5~eV for PrRu$_4$P$_{12}$ as 
seen in Figs.~1(b) and 2(b).      
This condition is actually met in the high pressure 
data of PrRu$_4$P$_{12}$, which has enabled us to derive 
its $\sigma(\omega)$ under pressure up to 14~GPa 
using the present method \cite{okamura}.    
%
%%%%%%%%%%%%%%%%%  FIG. 2  %%%%%%%%%%%%%%%
\begin{figure}
\begin{center}
\includegraphics[width=22pc]{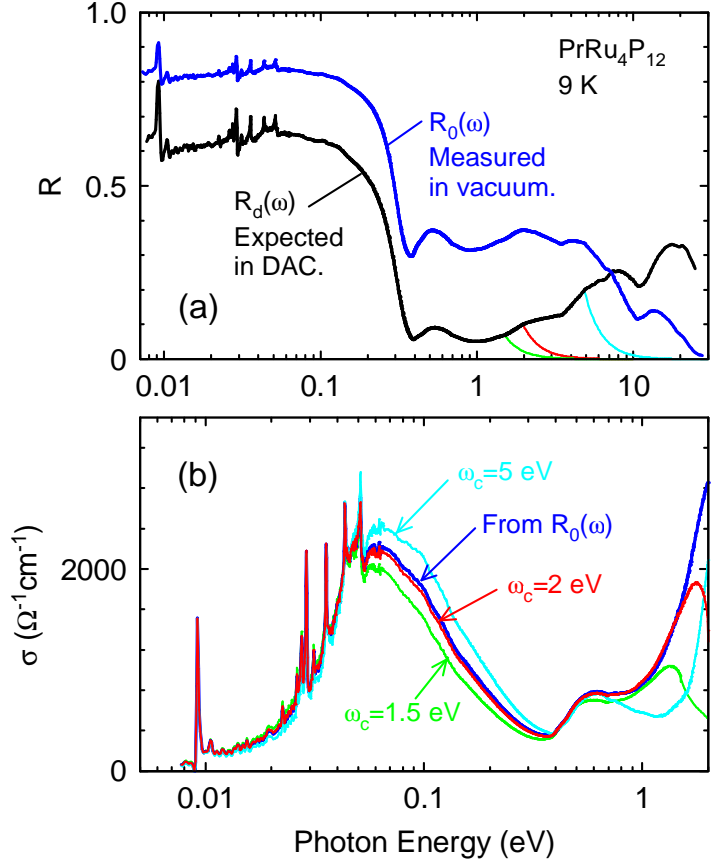}
\end{center}
\caption{(a) $R_0(\omega)$ is the reflectance spectrum 
of PrRu$_4$P$_{12}$ measured at 9~K in vacuum \cite{matunami}, 
and 
$R_{\rm d}(\omega)$ is that expected in a DAC calculated 
from $R_0(\omega)$ as described in the text.  The green, 
red, and light blue curves are $\omega^{-4}$ extrapolations 
with cutoff energies of $\omega_c$=1.5, 2, and 5~eV, 
respectively.  
(b) The optical conductivity ($\sigma$) obtained with KK 
analysis of $R_0(\omega)$ is compared with those obtained 
with KK analysis of $R_{\rm d}(\omega)$ with a cutoff at 
$\omega_c$=1.5, 2, and 5~eV, and $\omega^{-4}$ extrapolations 
above them.   
}
\end{figure}
%%%%%%%%%%%%%%%%%%%%%%%%%%%%%%%%%%%%%%%%%%
%%%%

We have also done similar simulations for other 
compounds, both metals and insulators, using 
actually measured data, and have 
obtained similar results.  Namely, when an appropriate 
cutoff and extrapolation are applied to $R_{\rm d}(\omega)$, 
the usual KK transform of Eq.~(6) gave $\sigma(\omega)$ 
spectra which agreed very well with those 
directly obtained from the wide range $R_0(\omega)$.   
A limitation of the present method is, as already 
mentioned above, it can give correct $\sigma(\omega)$ only 
below certain photon energy (1.5~eV in the case of 
PrRu$_4$P$_{12}$).   Hence this method is 
useful when the temperature and pressure dependences 
of $R_{\rm d}(\omega)$ is limited to below certain 
energy.    
In addition, to use the present method, it is required 
that $R_0(\omega)$ is known over a wide enough photon 
energy range, since $n(\omega)$ and $k(\omega)$ must 
be obtained from $R_0(\omega)$ with the usual KK analysis.     
While a mathematically rigorous justification of 
the proposed method is beyond the scope of this 
work, this method may be very useful as a simple analysis 
technique of reflectance spectra measured under high 
pressure with DAC.

\ack
This work has been done as a part of high pressure 
infrared studies of strongly correlated electron 
materials using synchrotron radiation at 
SPring-8, under the approval by JASRI 
(2009A0089 through 2011B0089).  
The $R_0(\omega)$ data used in Figs.~1 and 2 have been 
already published \cite{matunami}, in collaboration 
with M. Matsunami, L. Chen, 
M. Takimoto, T. Nanba, C. Sekine, and I. Shirotani.

\end{document}